\documentclass[aps,prl,twocolumn,superscriptaddress,nobibnotes,10pt]{revtex4-1}
\usepackage{scrextend}
\usepackage{graphicx}
\usepackage{amsmath}
\usepackage{ulem}

\usepackage{color}

\begin{document}


\title{Exploiting one-dimensional exciton-phonon coupling for tunable and efficient single-photon generation with a carbon nanotube.}

\author{A. Jeantet}
\author{Y. Chassagneux}
\author{T. Claude}
\author{P. Roussignol}
\affiliation{Laboratoire Pierre Aigrain, \'Ecole Normale Sup\'erieure, CNRS, Universit\'e Pierre et Marie Curie, Universit\'e Paris Diderot, PSL, Sorbonne Paris Cit\'e, Sorbonne Universit\'e, 24, rue Lhomond, F-75005 Paris, France}

\author{J.S. Lauret}
\affiliation{Laboratoire Aim\'e Cotton, CNRS, Univ.  Paris-Sud, ENS Cachan, Universit\'e Paris-Saclay, 91405 Orsay, France}
\author{J. Reichel}
\affiliation{Laboratoire Kastler Brossel, \'Ecole Normale Sup\'erieure, CNRS, Universit\'e Pierre et Marie Curie, 24 rue Lhomond, F-75005 Paris, France}

\author{C. Voisin}
\email{Corresponding author : christophe.voisin@lpa.ens.fr}
\affiliation{Laboratoire Pierre Aigrain, \'Ecole Normale Sup\'erieure, CNRS, Universit\'e Pierre et Marie Curie, Universit\'e Paris Diderot, PSL, Sorbonne Paris Cit\'e, Sorbonne Universit\'e, 24, rue Lhomond, F-75005 Paris, France}

\date{\today}

\begin{abstract}

Condensed-matter emitters offer enriched cavity quantum electrodynamical effects due to the coupling to external degrees of freedom. In the case of carbon nanotubes a very peculiar coupling between localized excitons and the one-dimensional acoustic phonon modes can be achieved, which gives rise to pronounced phonon wings in the luminescence spectrum. By coupling an individual nanotube to a tunable optical micro-cavity, we show that this peculiar exciton-phonon coupling is a valuable resource to enlarge the tuning range of the single-photon source while keeping an excellent exciton-photon coupling efficiency and spectral purity. Using the unique flexibility of our scanning fiber cavity, we are able to measure the efficiency spectrum of the very same nanotube in the Purcell regime for several mode volumes. Whereas this efficiency spectrum looks very much like the free-space luminescence spectrum when the Purcell factor is small (large mode volume), we show that the deformation of this spectrum at lower mode volumes can be traced back to the strength of the exciton-photon coupling. It shows an enhanced efficiency on the red wing that arises from the asymmetry of the incoherent energy exchange processes between the exciton and the cavity. This allows us to obtain a tuning range up to several hundred times the spectral width of the source. 

\end{abstract}


\maketitle


Light-matter interaction can be controlled at the ultimate level where a single quantum emitter interacts with a single photon by means of optical micro-cavities \cite{Vahala2003}. The small mode volumes and high quality factors achieved in such cavities provide a new means to tailor the optical properties of the emitter through the so-called cavity quantum electrodynamical (CQED) effects. In the weak coupling regime where the losses remain larger than the interaction strength between the emitter and the electromagnetic mode, it is possible to control the spontaneous emission rate of the emitter and to funnel efficiently the photons into a single photonic mode by reshaping the local density of states (Purcell effect \cite{Purcell1946}). This effect has been demonstrated with a number of solid-state emitters \cite{Aharonovich2016} and more recently with carbon nanotubes \cite{Miura2014, Jeantet2016, Pyatkov2016}. In this regime, a figure of merit of CQED effects is the Purcell factor $F_p$ defined as the ratio of the radiative emission rate into the cavity mode to the radiative rate in free-space. In addition, in the case of a broad emitter, the Purcell effect also allows to reshape the spectral properties of the emitter through the so-called cavity feeding effect. In particular for solid-state emitters, the environment induced broadening is a new resource that can be exploited to enhance the tuning range of the photon source, while keeping a narrow spectral output and large efficiency \cite{Auffeves2009}. 

This approach has been implemented in quantum dots where local electrostatic fluctuations and the three dimensional phonons of the matrix induce some level of pure dephasing of the excitonic line \cite{Portalupi2015}. Nevertheless, this effect remains limited to an energy window of the order of a few cavity line-widths ($\simeq 800\mu$eV). In carbon nanotubes at low temperature the situation is radically different because of the coupling of localized excitons to the truly one-dimensional acoustic phonon bath. This gives rise to a diverging effective exciton-phonon coupling for low-energy vibrational modes resulting in phonon wings extending over several meV even though the zero-phonon line (ZPL) is typically 10 to 100 times narrower \cite{Galland2008, Vialla2014, Ardizzone2015, Sarpkaya2015}. As first noted by Galland \textit{et al.} the strong asymmetry of the wings is the signature of non-Markovian decoherence processes due to the one-dimensional confinement of acoustic phonons \cite{Galland2008}. In addition, this dephasing is to some extent adjustable by means of proper nanotube/matrix coupling resulting in variable exciton and phonon localization lengths \cite{Vialla2014}. When the nanotube is coupled to a tunable micro-cavity, this spectral broadening enlarges considerably the tuning range of the photon source while keeping excellent anti-bunching properties \cite{Jeantet2016}. 

In this work, we show that this peculiar exciton-phonon coupling results in an asymmetric incoherent energy exchange between the cavity mode and the emitter. This enables an enhanced cavity feeding effect on the red wing, possibly by-passing the intrinsic limit for the efficiency of the source. We investigate this effect experimentally by scanning continuously the working frequency of a fiber micro-cavity coupled to a nanotube, thereby measuring the single-photon generation spectral efficiency. Beside a globally enhanced brightness, we show that the spectral efficiency of the source at lower mode volumes does not map out the free-space PL spectrum of the emitter anymore. In particular, the efficiency of the source on the red phonon wing is more enhanced than at the ZPL. This effect is interpreted quantitatively in the framework of the Jaynes-Cummings model where the exciton-phonon coupling is described in the spin-boson approximation \cite{Krummheuer2002,Leggett1987,Wilson-Rae2002,theorypaper} with an effective phonon mode taking into account the coupling to all acoustic branches \cite{Nguyen2011a}. 

The experimental setup was described in Ref. \cite{Jeantet2016}. In brief, PFO coated CoMoCat nanotubes embedded in a 120~nm thick layer of polystyrene \cite{Ai2011} are deposited on a flat dielectric mirror. All the measurements are conducted at 15~K. The individual nanotubes are selected by means of a micro-photoluminescence (PL) setup using a cw excitation wavelength near 800~nm.  After a full free-space characterization the nanotube is inserted into a micro-cavity by approaching the top mirror (radius of curvature of $\sim 10 \mu$m) which is engineered at the apex of an optical fiber by laser ablation \cite{Hunger2010}. The fiber is mounted on a three-dimensional piezo stage in order to adjust the spatial mode matching and the cavity length. The luminescence collected from the cavity or from the PL microscope (free-space emission) is dispersed in a 500~mm spectrometer and detected with a nitrogen-cooled CCD camera.

\begin{figure}
\includegraphics[width=8.5cm]{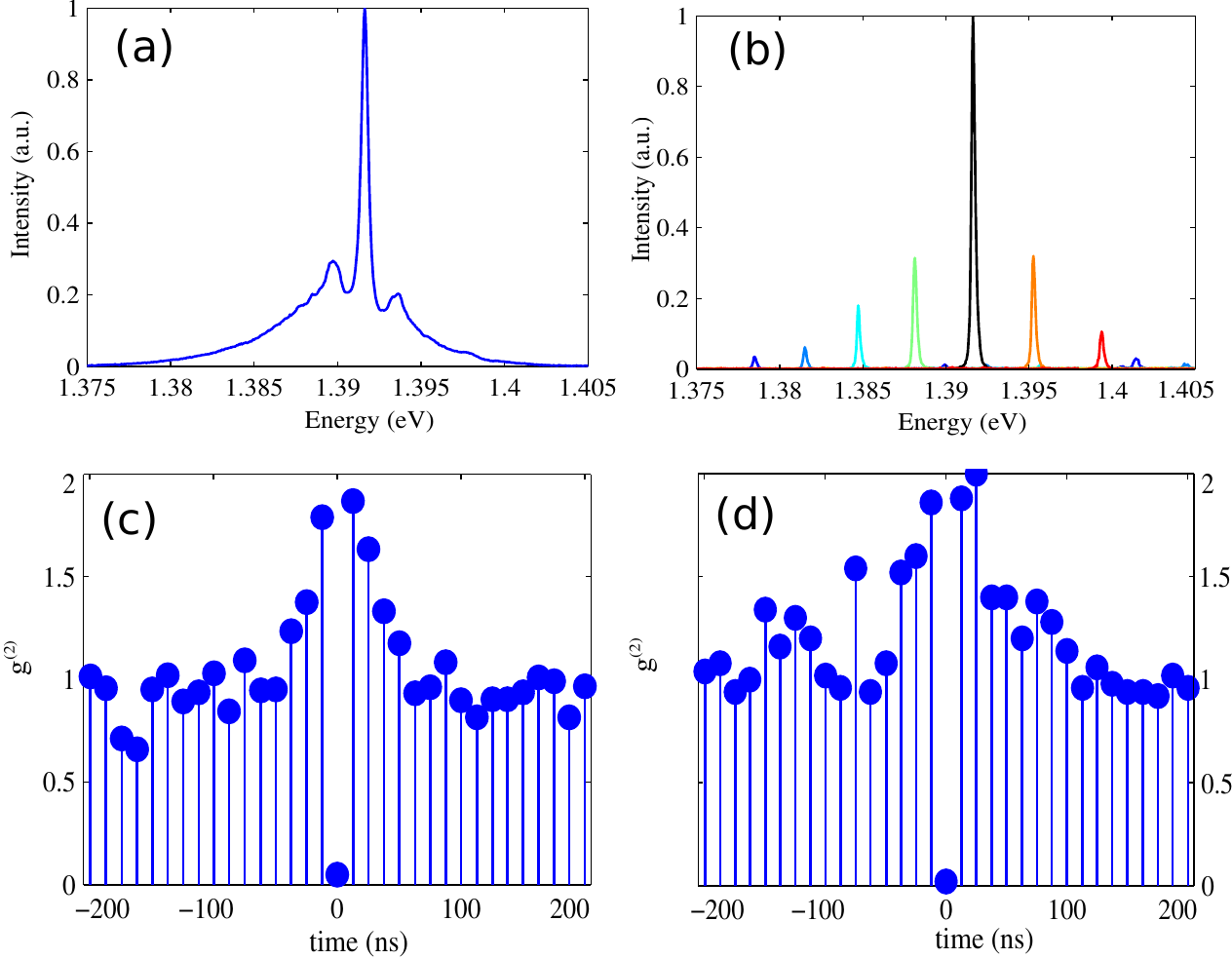}
\caption{(a) Free-space PL spectrum of a single carbon nanotube recorded at 15~K for a cw excitation at $\lambda=$800~nm. (b) Spectra of the single-photon source obtained with this nanotube for several cavity detunings showing the high spectral purity and high efficiency of the single-photon source on a broad tuning range. The excitation density is kept constant for all the spectra. (c) and (d) Intensity correlation functions $g^{(2)}(t)$ measured under pulsed excitation for a nanotube coupled to a cavity in resonance with the ZPL (c) or in resonance with the red phonon wing at a detuning of 5~meV (d). (For technical reasons, these data were not recorded on the nanotube studied in the rest of the paper and correspond to the data published in \cite{Jeantet2016}.)\label{fig:spectres_et_g2} }
\end{figure}

The free-space luminescence spectrum of a carbon nanotube at 15~K is presented in Fig.~\ref{fig:spectres_et_g2}~(a). 
Its characteristic profile consists of a sharp ZPL together with a low intensity blue wing and a higher intensity red wing. This profile is well understood in the framework of one-dimensional acoustic phonon side bands of a localized exciton \cite{Vialla2014, Ardizzone2015, Sarpkaya2015}. 
As a consequence of phonon confinement in PFO-wrapped nanotubes, the ZPL becomes largely decoupled from the phonon wings \cite{Vialla2014, Sarpkaya2015}. 

%

When phonon side-bands represent a sizable fraction of the transition oscillator strength, the CQED effects modify the recombination dynamics both for the ZPL and for the sidebands, which makes it possible to tune the source over a broad spectral range with a large efficiency (Fig.~\ref{fig:spectres_et_g2}~(b)). We checked that the anti-bunching level remains below $0.03\pm0.02$ both at the ZPL frequency and on the phonon wings (Fig.~\ref{fig:spectres_et_g2}~(c-d)). In addition this anti-bunching is very robust to an increased pumping rate since it remains below a few percents up to a pumping intensity of the order of the saturation intensity.

\begin{figure}[h]
\includegraphics[width=8.5cm]{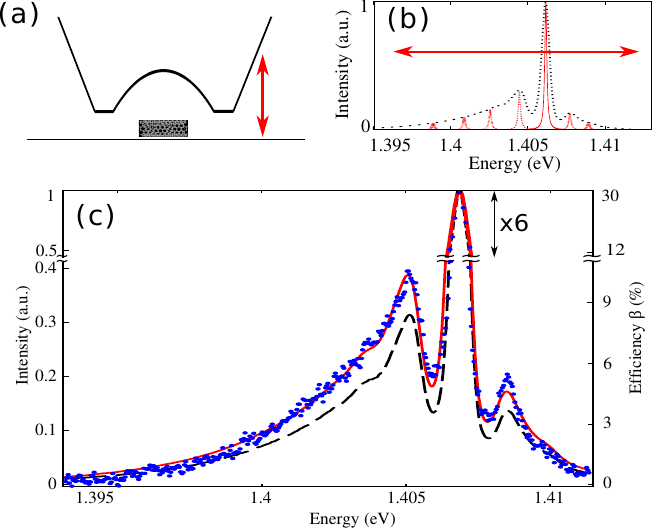}
\caption{ (a) A nanotube is in the cavity formed by a planar mirror and a concave mirror engineered at the tip of an optical fiber. The cavity length oscillates at frequency $\Omega$ to cover the full nanotube spectrum. (b) Sketch of the spectrum of the source for selected detunings (e.g. solid red line when the cavity is in resonance with the ZPL). The dotted line represents the efficiency spectrum as obtained from the envelope of the spectra when the detuning is scanned continuously (see main text). (c) Experimental efficiency spectrum (blue dots) for the lowest mode volume together with the normalized free-space spectrum (dashed black) of the same nanotube and a fit to the data (solid red) according to eq.(~\ref{eq:coupling_efficiency}). Note that the data from all the figures 2-5 were taken on this same nanotube, as well as the values discussed in the main text. \label{fig:cavity-scan}}
\end{figure}

In order to quantify the strength of the exciton-photon coupling in the cavity feeding regime, we traced back the coupling strength $\hbar g$ to the dependence of the efficiency of the photon source on the detuning $\hbar \omega_{cav}-\hbar \omega_{ZPL}$.  This coupling $g$ (equal to half the vacuum Rabi splitting) corresponds to the interaction energy between the transition dipole and the electrical field of the zero-point quantum fluctuations of the optical mode. The single-photon source efficiency $\beta(f)$ (where $f$ is the optical frequency) is defined as the probability for a photon to be emitted by the cavity for each single excitation launched into the nanotube. For a given excitation density, this efficiency is proportional to the output intensity. In fact, for a fixed cavity length, the output of the source $I(f)$ is a narrow line as shown in Fig.~\ref{fig:spectres_et_g2}~(b) which simply reads $I(f) \propto S^{cav}_{f_c} (f) \cdot \beta(f_c) $, where $S^{cav}$ is the normalized Lorentzian profile of the cavity centered at $f_c$. In order to get the variation of the efficiency with the detuning, we scan the cavity length continuously at a frequency $\Omega=25$~Hz and record the time-average spectrum of the output (Fig.~\ref{fig:cavity-scan}~(a)). The cavity scanning range spans a window slightly broader than the nanotube emission spectrum (Fig.~\ref{fig:cavity-scan}~(b)) so that the dwell time of the cavity is constant all over the spectrum. A strong enhancement of the signal is observed at resonance with the ZPL, whereas the intensity remains sizable when the cavity probes the phonon side bands (Fig.~\ref{fig:cavity-scan}~(c)). Note that the full scanning window ($\simeq20$~meV) corresponds to a relative change in the cavity length of $1.4\%$ and thus to a change in the mode volume of about $2\%$, which is negligible (see SI in \cite{Jeantet2016}). 

\begin{figure}[h]
\includegraphics[width=8.5cm]{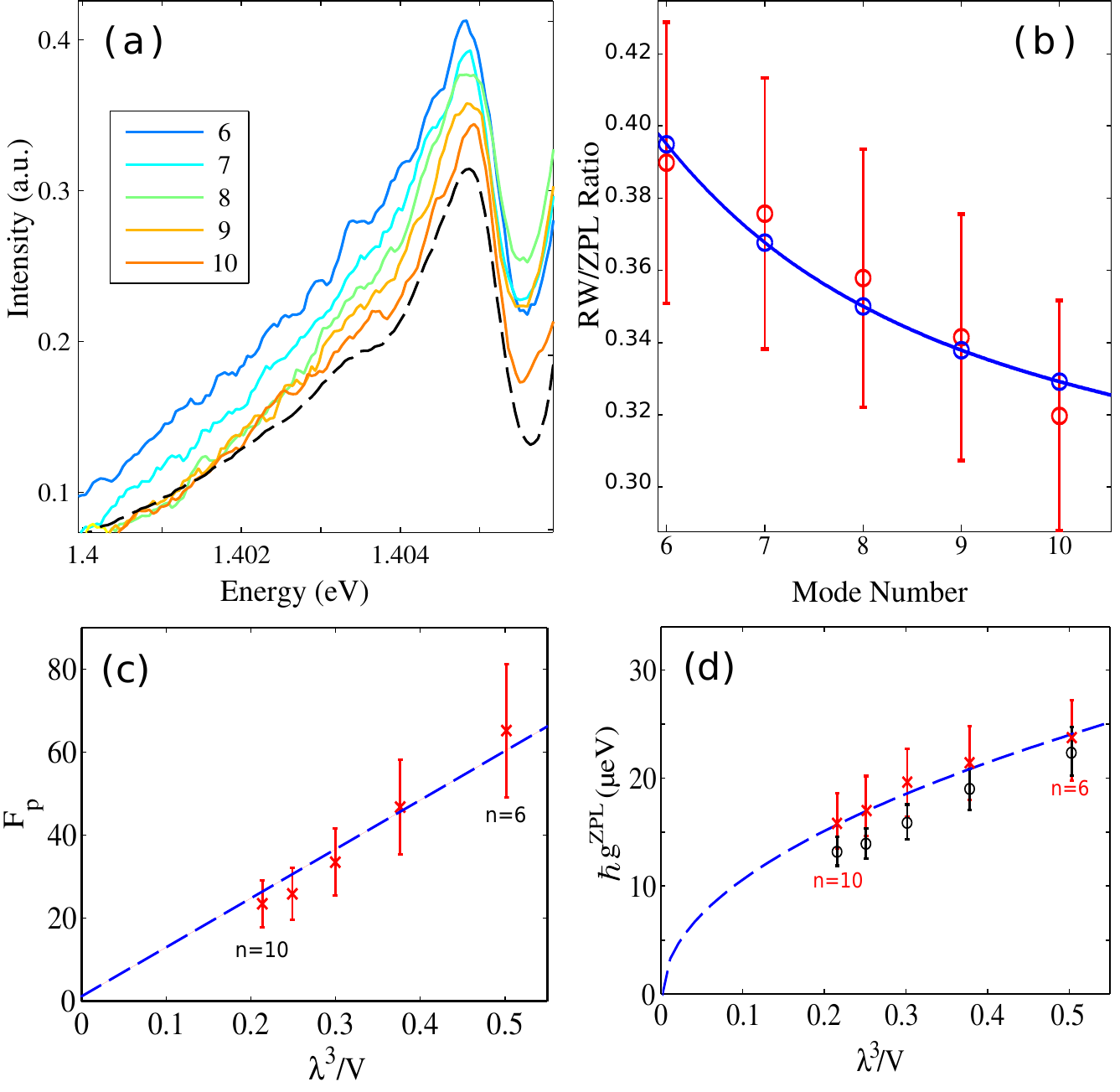}
\caption{ (a) Evolution of the spectral efficiency of the same emitter for selected mode volumes labeled by longitudinal mode number (zoom on the red phonon wing). (b) Ratio of the intensity at the maximum of the red wing to the maximum of the ZPL as a function of the longitudinal mode number (red dots). Theoretical values (blue open dots) as obtained from eq.~(\ref{eq:coupling_efficiency}) and the data of panel (d). The solid blue line is a guide to the eye. (c) Purcell factor of an individual nanotube coupled to a micro-cavity tuned at the ZPL frequency for several cavity lengths, extracted from lifetime measurements (eq. (\ref{eq:Fp_vs_lifetime})). The blue dashed line shows the 1/V behavior. (d) Coupling factor $g_{ZPL}$ for different mode volumes, extracted from lifetime measurements (red crosses) and from the fit of the efficiency spectrum according to eq. (\ref{eq:coupling_efficiency}) (black open dots). The dashed blue line shows the $\sqrt{1/V}$ behavior. The mode volume is deduced from the cavity free spectral range (FSR) \cite{Jeantet2016}. \label{fig:spectres}}
\end{figure}

This efficiency spectrum can also be measured for several values of the mode volume by changing the length of the cavity by steps of $\lambda/2$, \textit{i.e.} by changing the longitudinal mode order. Overall, the mode volume can be varied by at least of factor of 2. Basically, the output intensity increases when decreasing the mode volume as expected from the Purcell effect (Fig.~\ref{fig:spectres}~(a)). However, the spectra are not homothetic to each other and the intensity of the phonon wings grows at a faster pace than that of the ZPL when decreasing the mode volume (Fig.\ref{fig:spectres} (b)).

 This behavior can be captured quantitatively by a generalized Jaynes Cummings model where the coupling to the phonon bath is treated in the spin-boson framework within the NIBA approximation \cite{theorypaper}. In this approach, an elementary excitation is launched in the exciton population and the efficiency of the source is calculated as the time-integrated output of the cavity mode. The outcome of the model is that the efficiency of the source reads :
 
 \begin{equation}
 \beta(f_{cav})=\frac{g^2\kappa \tilde{S}^{\mathrm{em}}(f_{cav})}{g^2 \kappa \tilde{S}^{\mathrm{em}}(f_{cav}) + g^2 \gamma \tilde{S}^{\mathrm{abs}}(f_{cav}) + \kappa \gamma}
 \label{eq:coupling_efficiency}
\end{equation}

where $\kappa$ stands for the cavity decay rate, $\gamma$ for the exciton free-space recombination rate, $\tilde{S}^{\mathrm{em}}(f)$ and $\tilde{S}^{\mathrm{abs}}(f)$ are normalized ($\int S(f) df \equiv 1$) emission and absorption free-space spectra convoluted by the normalized Lorentzian cavity mode profile. Note that when the cavity mode is much narrower than the ZPL, the convolution can be omitted and $\tilde{S}^{\mathrm{em}}(f)$ represents the probability for the free-space radiative recombination to occur within a frequency band d$f$ around $f$ when one exciton is launched in the nanotube. 

\begin{figure}[h]
\includegraphics[width=8.5cm]{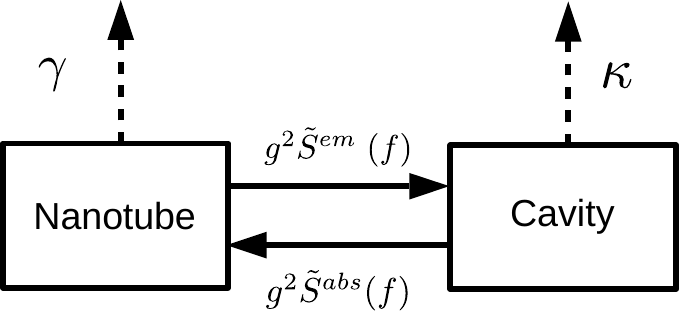}
\caption{Heuristic model of the dynamics of the exciton/cavity/phonon system based on incoherent energy exchanges between the emitter and the cavity. \label{fig:theory-sketch} }
\end{figure}

Although this formula is established in a rigorous way within the spin-boson model \cite{theorypaper}, it is enlightening to infer it in a heuristic way by considering incoherent energy exchanges between the emitter and the cavity mode (Fig.~\ref{fig:theory-sketch}). In this picture, one can show that when the cavity is detuned, the energy exchange rate becomes $g^2 \tilde{S}^{em}$ for the emission process whereas it becomes $g^2 \tilde{S}^{abs}$ for the absorption process (SI). As a consequence, the total decay rate of the emitter becomes $\gamma + g^2 \tilde{S}^{em}$ whereas the total decay rate of the cavity mode becomes $\kappa + g^2 \tilde{S}^{abs}$. Hence, the probability for a photon in the cavity mode to reach the detector reads :
\begin{equation}
 p_{cav \rightarrow det}= \kappa/[\kappa + g^2 \tilde{S}^{abs}]
\end{equation}

the probability for the photon to be reabsorbed by the emitter reads :
\begin{equation}
 p_{cav \rightarrow NT}= g^2 \tilde{S}^{abs}/[\kappa + g^2 \tilde{S}^{abs}]
\end{equation}

and the probability for the excited emitter to emit into the cavity mode reads :
\begin{equation}
 p_{NT \rightarrow cav}= g^2 \tilde{S}^{em}/[\gamma + g^2 \tilde{S}^{em}]
\end{equation}

Finally, the total probability for a photon to be emitted by the source when a single excitation is launched in the emitter (which is the above defined efficiency of the source) reads :
 
 \begin{multline}
 \beta = p_{NT \rightarrow det}= p_{NT \rightarrow cav}[1+p_{cav \rightarrow NT}p_{NT \rightarrow cav}+...\\
 +(p_{cav \rightarrow NT}p_{NT \rightarrow cav})^n]p_{cav \rightarrow det} 
 \end{multline}
 
which yields eq.~(\ref{eq:coupling_efficiency}). This heuristic approach shows that the combined effect of the exciton-phonon and exciton-photon couplings boils down to asymmetric energy exchanges between the cavity and the nanotube. In addition, the full derivation shows that this conclusion remains valid even when the coupling $g$ becomes larger than the loss rates $\gamma$ and $\kappa$ (strong coupling regime in a two-level picture) \cite{theorypaper}.
 
The emission spectrum $\tilde{S}^{em}$ can be directly obtained from the measurement of the free-space photoluminescence spectrum. The absorption spectrum of the nanotube can be deduced from a mirror symmetry with respect to the ZPL as is well known for emitters showing Huang-Rhys phonon coupling profiles \cite{Henderson1989}.

Figure~\ref{fig:cavity-scan}~(c) shows the efficiency spectrum measured for the smallest mode volume of $\sim 2 \lambda^3$ obtained for a cavity length of $6 \lambda/2$ together with the free-space emission spectrum normalized to the same ZPL intensity.  Obviously, the relative efficiency of the single-photon source at the phonon wings  is boosted by the cavity as compared to the intensity at the ZPL frequency. This behavior is quantitatively reproduced by the  model as shown by the red line which is a fit according to eq.~(\ref{eq:coupling_efficiency}) with a single free parameter, the coupling strength $g$ (all the other parameters : loss rates, emission profile,... are deduced from independent measurements, see SI). 
In addition, the ratio of the efficiency of the source at the red phonon wing maximum over the one at the ZPL shows a dependence as a function of the cavity length (Fig.~\ref{fig:spectres}~(b)) which is well accounted for by eq.~(\ref{eq:coupling_efficiency}), using the fact that $g^2 \propto 1/V$. We obtain values ranging from $\hbar g=23\pm4 \mu$eV to $\hbar g=40\pm4 \mu$eV for the largest and smallest mode volumes respectively (obtained for longitudinal mode orders ranging from 6 to 10).

Interestingly, these values of $g$ can be benchmarked to those deduced from time-resolved photo-luminescence. In this case, we focus on measurements done at the ZPL. Using pulsed excitation and a fast photon counting module, we obtain the life-time $\tau$ of the exciton both in free-space and in the cavity for several mode volumes (SI). The change of lifetime $\Delta \tau$ gives access to the effective Purcell factor $F_p$ provided that the PL quantum yield $\eta$ of the emitter is known \cite{Faraon2011} : 

\begin{equation}
  F_p= \frac{\Delta \tau}{\eta \tau}
  \label{eq:Fp_vs_lifetime}
\end{equation}

In the present case $\eta$ represents the effective quantum yield of the ZPL which includes the Debye-Waller factor. It is obtained from saturation measurements in pulsed excitation \cite{Jeantet2016}. As expected, the Purcell factors we deduce from this method follow a $1/V$ law (Fig.~\ref{fig:spectres}~(c)), where $V$ is the mode volume, in agreement with the Purcell formula $F_p=3/(4\pi^2) (\lambda/n)^3 Q_e/V$ where $\lambda$ is the emission wavelength, $n$ is the optical index and $Q_e$ is an effective quality factor such as $Q_e^{-1}=Q_{cav}^{-1}+Q_{ZPL}^{-1}$ \cite{Auffeves2010}.
Notably, these values of $F_p$, up to $F_p=60$ are much larger than our previous data (of the order of 5 to 10) \cite{Jeantet2016}. This arises from the much reduced radius of curvature of the top mirror leading to a mode-volume reduction of the order of 5 in line with the observed increase of $F_p$. Such high values of $F_p$ are particularly valuable for a dim emitter like carbon nanotubes (that show an intrinsic quantum yield $\eta$ of the order of 1\%) because it directly translates into an enhanced brightness of the source even for a pumping intensity well below the saturation. Here, we infer that the effective quantum yield of the nanotube coupled to the cavity amounts up to $\eta F_p/(1+\eta F_p) \sim 40\% \pm 5$, which is in qualitative agreement with the values deduced from the fitting of the efficiency spectrum with eq. (\ref{eq:coupling_efficiency}). Given the short life-time of nanotubes (100~ps or below) and the nearly perfect extraction efficiency, we anticipate that the single-photon generation rate could reach up to several GHz.

This cavity induced reduction of life-time is a direct consequence of the strength of the exciton-photon coupling $g_{ZPL}$. In a simple Lorentzian picture, which is reasonably applicable for the ZPL, the coupling $g_{ZPL}$ can be traced back to the change of lifetime through $g_{ZPL}=1/2\sqrt{(1/\tau_{cav}-1/\tau_{\mathrm{fs}})\omega_{ZPL}/Q_e}$, where $\tau_{cav}$ (resp. $\tau_{\mathrm{fs}}$) stands for the lifetime of the emitter coupled to the cavity tuned at the angular frequency $\omega_{ZPL}$ (resp. the free-space lifetime). Interestingly, $g_{ZPL}$ can be measured experimentally with higher accuracy than $F_p$ since it does not rely on any calibration of the setup detection efficiency (needed to assess the quantum efficiency used in (\ref{eq:Fp_vs_lifetime})). Finally, the coupling deduced from this life-time measurements can be related to $g$ deduced from the fitting of the efficiency spectra using $g_{ZPL}=g*\sqrt{DW}$, where $DW=0.4$ is the Debye-Waller factor (SI). The two approaches agree within the error bars as shown in Fig.~\ref{fig:spectres}~(d).

 \begin{figure}[h]
\includegraphics[width=8.5cm]{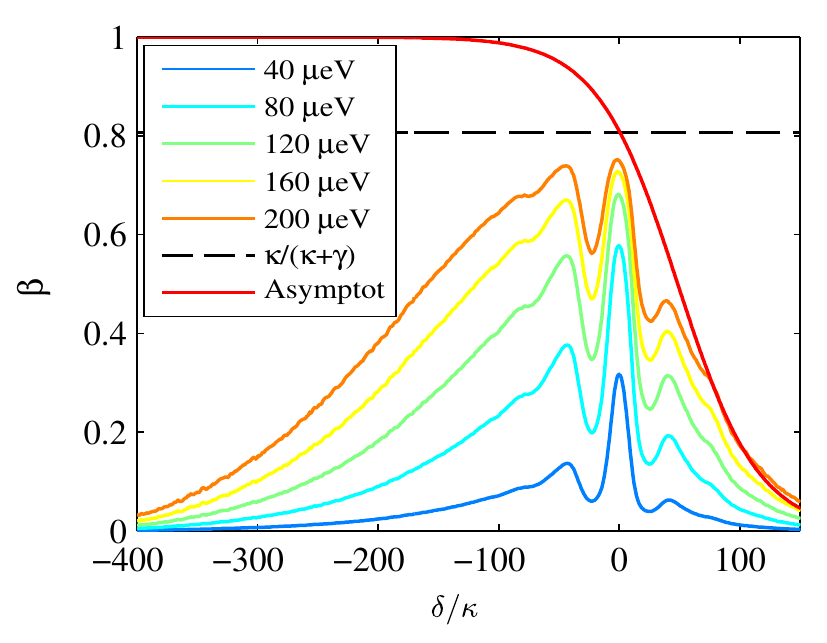}
\caption{Simulated evolution of the efficiency (for different values of the coupling parameter $g$) as a function of the cavity detuning $\delta=\omega-\omega_{ZPL}$ expressed in units of the cavity linewidth $\kappa$. This representation shows readily the remarkable tuning capabilities of the source. The red line corresponds to the asymptotic efficiency $\beta^\infty$ while the dashed black one corresponds to the intrinsic limit for markovian decoherence. \label{fig:theory-spectres}}
\end{figure}

It is interesting to examine the asymptotic behavior of the system when the coupling strength $g$ is increased (eq.~(\ref{eq:coupling_efficiency})). The efficiency spectrum becomes more and more distorted, the red phonon wing becoming eventually higher than the ZPL (Fig.~\ref{fig:theory-spectres}). This striking effect arises from the asymmetry between the absorption and emission processes on the red phonon wing at low temperature, the former being much less likely than the latter on the red side. Thus, the nanotube cavity coupling becomes favorably asymmetric with enhanced emission probability of the nanotube into the cavity mode and reduced reabsorption probability. This allows to bypass the natural limit $\kappa/(\kappa+\gamma)$ observed at the ZPL where both processes are equiprobable. For $\hbar g$ values of the order of 40~$\mu$eV and for nanotubes showing a ZPL width of the order of 900~$\mu$eV the tuning range of the source can already reach up to 10~meV (2.5~THz). Using a stabilized cavity, the source width (itself set by the cavity mode width) can be reduced down to below 10~GHz \cite{Hunger2010}. Thus, the tuning range represents up to several hundred times the spectral width of the source, opening the way to wide-range multiplexing in quantum communications. This is in contrast to the plasmonic approach where the large tuning range is obtained at the expense of the brightness \cite{Zakharko2016}.

In conclusion we have shown that the peculiar exciton-phonon coupling in carbon nanotubes is a valuable resource to expand the bandwidth of a single-photon source. The most original feature of this system is the asymmetry in the energy exchange processes that allows to bypass the intrinsic limit for the efficiency of the source and ultimately to reach a close to 100\% yield, provided that the coupling is strong enough. This could be achieved in future integrated devices using photonic crystal resonators that have much reduced mode volumes. In addition, the use of carbon nanotubes of diameters of the order of 1.1~nm or of functionalized carbon nanotubes would bring the single-photon source in the telecom bands \cite{Ma2015a}. The indistinguishability of our source is currently estimated to 0.25 (see SI), but could be sensibly increased by tuning
the coupling and finesse of the cavity \cite{Grange2015} thus paving the way to the tunable generation of indistinguishable single photons using carbon nanotubes.

\begin{acknowledgments}
This work was supported by the C'Nano IdF grant \textit{"ECOQ"} and the ANR grant \textit{``NC2''}. JSL is partly funded by ``Institut Universitaire de France''.  We thank A. Leclercq for the mechanical fabrication, N. Izard for helping in sample preparation. \\
\end{acknowledgments}

\bibliography{NTcav-spectrum}

\begin{thebibliography}{25}%
\makeatletter
\providecommand \@ifxundefined [1]{%
 \@ifx{#1\undefined}
}%
\providecommand \@ifnum [1]{%
 \ifnum #1\expandafter \@firstoftwo
 \else \expandafter \@secondoftwo
 \fi
}%
\providecommand \@ifx [1]{%
 \ifx #1\expandafter \@firstoftwo
 \else \expandafter \@secondoftwo
 \fi
}%
\providecommand \natexlab [1]{#1}%
\providecommand \enquote  [1]{``#1''}%
\providecommand \bibnamefont  [1]{#1}%
\providecommand \bibfnamefont [1]{#1}%
\providecommand \citenamefont [1]{#1}%
\providecommand \href@noop [0]{\@secondoftwo}%
\providecommand \href [0]{\begingroup \@sanitize@url \@href}%
\providecommand \@href[1]{\@@startlink{#1}\@@href}%
\providecommand \@@href[1]{\endgroup#1\@@endlink}%
\providecommand \@sanitize@url [0]{\catcode `\\12\catcode `\$12\catcode
  `\&12\catcode `\#12\catcode `\^12\catcode `\_12\catcode `\%12\relax}%
\providecommand \@@startlink[1]{}%
\providecommand \@@endlink[0]{}%
\providecommand \url  [0]{\begingroup\@sanitize@url \@url }%
\providecommand \@url [1]{\endgroup\@href {#1}{\urlprefix }}%
\providecommand \urlprefix  [0]{URL }%
\providecommand \Eprint [0]{\href }%
\providecommand \doibase [0]{http://dx.doi.org/}%
\providecommand \selectlanguage [0]{\@gobble}%
\providecommand \bibinfo  [0]{\@secondoftwo}%
\providecommand \bibfield  [0]{\@secondoftwo}%
\providecommand \translation [1]{[#1]}%
\providecommand \BibitemOpen [0]{}%
\providecommand \bibitemStop [0]{}%
\providecommand \bibitemNoStop [0]{.\EOS\space}%
\providecommand \EOS [0]{\spacefactor3000\relax}%
\providecommand \BibitemShut  [1]{\csname bibitem#1\endcsname}%
\let\auto@bib@innerbib\@empty
\bibitem [{\citenamefont {Vahala}(2003)}]{Vahala2003}%
  \BibitemOpen
  \bibfield  {author} {\bibinfo {author} {\bibfnamefont {K.~J.}\ \bibnamefont
  {Vahala}},\ }\href@noop {} {\bibfield  {journal} {\bibinfo  {journal}
  {Nature}\ }\textbf {\bibinfo {volume} {424}},\ \bibinfo {pages} {839}
  (\bibinfo {year} {2003})}\BibitemShut {NoStop}%
\bibitem [{\citenamefont {Purcell}(1946)}]{Purcell1946}%
  \BibitemOpen
  \bibfield  {author} {\bibinfo {author} {\bibfnamefont {E.~M.}\ \bibnamefont
  {Purcell}},\ }\href@noop {} {\bibfield  {journal} {\bibinfo  {journal} {Phys.
  Rev.}\ }\textbf {\bibinfo {volume} {69}},\ \bibinfo {pages} {681} (\bibinfo
  {year} {1946})},\ \bibinfo {note} {abstract B10, page 681}\BibitemShut
  {NoStop}%
\bibitem [{\citenamefont {Aharonovich}\ \emph {et~al.}(2016)\citenamefont
  {Aharonovich}, \citenamefont {Englund},\ and\ \citenamefont
  {Toth}}]{Aharonovich2016}%
  \BibitemOpen
  \bibfield  {author} {\bibinfo {author} {\bibfnamefont {I.}~\bibnamefont
  {Aharonovich}}, \bibinfo {author} {\bibfnamefont {D.}~\bibnamefont
  {Englund}}, \ and\ \bibinfo {author} {\bibfnamefont {M.}~\bibnamefont
  {Toth}},\ }\href {\doibase 10.1038/nphoton.2016.186} {\bibfield  {journal}
  {\bibinfo  {journal} {Nature Photonics}\ }\textbf {\bibinfo {volume} {10}},\
  \bibinfo {pages} {631–641} (\bibinfo {year} {2016})}\BibitemShut {NoStop}%
\bibitem [{\citenamefont {Miura}\ \emph {et~al.}(2014)\citenamefont {Miura},
  \citenamefont {Imamura}, \citenamefont {Ohta}, \citenamefont {Ishii},
  \citenamefont {Liu}, \citenamefont {Shimada}, \citenamefont {Iwamoto},
  \citenamefont {Arakawa},\ and\ \citenamefont {Kato}}]{Miura2014}%
  \BibitemOpen
  \bibfield  {author} {\bibinfo {author} {\bibfnamefont {R.}~\bibnamefont
  {Miura}}, \bibinfo {author} {\bibfnamefont {S.}~\bibnamefont {Imamura}},
  \bibinfo {author} {\bibfnamefont {R.}~\bibnamefont {Ohta}}, \bibinfo {author}
  {\bibfnamefont {A.}~\bibnamefont {Ishii}}, \bibinfo {author} {\bibfnamefont
  {X.}~\bibnamefont {Liu}}, \bibinfo {author} {\bibfnamefont {T.}~\bibnamefont
  {Shimada}}, \bibinfo {author} {\bibfnamefont {S.}~\bibnamefont {Iwamoto}},
  \bibinfo {author} {\bibfnamefont {Y.}~\bibnamefont {Arakawa}}, \ and\
  \bibinfo {author} {\bibfnamefont {Y.~K.}\ \bibnamefont {Kato}},\ }\href
  {http://dx.doi.org/10.1038/ncomms6580} {\bibfield  {journal} {\bibinfo
  {journal} {Nat Commun}\ }\textbf {\bibinfo {volume} {5}},\  (\bibinfo {year}
  {2014})}\BibitemShut {NoStop}%
\bibitem [{\citenamefont {Jeantet}\ \emph {et~al.}(2016)\citenamefont
  {Jeantet}, \citenamefont {Chassagneux}, \citenamefont {Raynaud},
  \citenamefont {Roussignol}, \citenamefont {Lauret}, \citenamefont {Besga},
  \citenamefont {Est\`eve}, \citenamefont {Reichel},\ and\ \citenamefont
  {Voisin}}]{Jeantet2016}%
  \BibitemOpen
  \bibfield  {author} {\bibinfo {author} {\bibfnamefont {A.}~\bibnamefont
  {Jeantet}}, \bibinfo {author} {\bibfnamefont {Y.}~\bibnamefont
  {Chassagneux}}, \bibinfo {author} {\bibfnamefont {C.}~\bibnamefont
  {Raynaud}}, \bibinfo {author} {\bibfnamefont {P.}~\bibnamefont {Roussignol}},
  \bibinfo {author} {\bibfnamefont {J.}~\bibnamefont {Lauret}}, \bibinfo
  {author} {\bibfnamefont {B.}~\bibnamefont {Besga}}, \bibinfo {author}
  {\bibfnamefont {J.}~\bibnamefont {Est\`eve}}, \bibinfo {author}
  {\bibfnamefont {J.}~\bibnamefont {Reichel}}, \ and\ \bibinfo {author}
  {\bibfnamefont {C.}~\bibnamefont {Voisin}},\ }\href {\doibase
  10.1103/physrevlett.116.247402} {\bibfield  {journal} {\bibinfo  {journal}
  {Physical Review Letters}\ }\textbf {\bibinfo {volume} {116}},\ \bibinfo
  {pages} {247402} (\bibinfo {year} {2016})}\BibitemShut {NoStop}%
\bibitem [{\citenamefont {Pyatkov}\ \emph {et~al.}(2016)\citenamefont
  {Pyatkov}, \citenamefont {F\"utterling}, \citenamefont {Khasminskaya},
  \citenamefont {Flavel}, \citenamefont {Hennrich}, \citenamefont {Kappes},
  \citenamefont {Krupke},\ and\ \citenamefont {Pernice}}]{Pyatkov2016}%
  \BibitemOpen
  \bibfield  {author} {\bibinfo {author} {\bibfnamefont {F.}~\bibnamefont
  {Pyatkov}}, \bibinfo {author} {\bibfnamefont {V.}~\bibnamefont
  {F\"utterling}}, \bibinfo {author} {\bibfnamefont {S.}~\bibnamefont
  {Khasminskaya}}, \bibinfo {author} {\bibfnamefont {B.~S.}\ \bibnamefont
  {Flavel}}, \bibinfo {author} {\bibfnamefont {F.}~\bibnamefont {Hennrich}},
  \bibinfo {author} {\bibfnamefont {M.~M.}\ \bibnamefont {Kappes}}, \bibinfo
  {author} {\bibfnamefont {R.}~\bibnamefont {Krupke}}, \ and\ \bibinfo {author}
  {\bibfnamefont {W.~H.~P.}\ \bibnamefont {Pernice}},\ }\href {\doibase
  10.1038/nphoton.2016.70} {\bibfield  {journal} {\bibinfo  {journal} {Nature
  Photon}\ }\textbf {\bibinfo {volume} {10}},\ \bibinfo {pages} {420–427}
  (\bibinfo {year} {2016})}\BibitemShut {NoStop}%
\bibitem [{\citenamefont {Auff\`eves}\ \emph {et~al.}(2009)\citenamefont
  {Auff\`eves}, \citenamefont {G\'erard},\ and\ \citenamefont
  {Poizat}}]{Auffeves2009}%
  \BibitemOpen
  \bibfield  {author} {\bibinfo {author} {\bibfnamefont {A.}~\bibnamefont
  {Auff\`eves}}, \bibinfo {author} {\bibfnamefont {J.-M.}\ \bibnamefont
  {G\'erard}}, \ and\ \bibinfo {author} {\bibfnamefont {J.-P.}\ \bibnamefont
  {Poizat}},\ }\href {http://link.aps.org/doi/10.1103/PhysRevA.79.053838}
  {\bibfield  {journal} {\bibinfo  {journal} {Phys. Rev. A}\ }\textbf {\bibinfo
  {volume} {79}},\ \bibinfo {pages} {053838} (\bibinfo {year}
  {2009})}\BibitemShut {NoStop}%
\bibitem [{\citenamefont {Portalupi}\ \emph {et~al.}(2015)\citenamefont
  {Portalupi}, \citenamefont {Hornecker}, \citenamefont {Giesz}, \citenamefont
  {Grange}, \citenamefont {Lema\^itre}, \citenamefont {Demory}, \citenamefont
  {Sagnes}, \citenamefont {Lanzillotti-Kimura}, \citenamefont {Lanco},
  \citenamefont {Auff\`eves},\ and\ \citenamefont {Senellart}}]{Portalupi2015}%
  \BibitemOpen
  \bibfield  {author} {\bibinfo {author} {\bibfnamefont {S.~L.}\ \bibnamefont
  {Portalupi}}, \bibinfo {author} {\bibfnamefont {G.}~\bibnamefont
  {Hornecker}}, \bibinfo {author} {\bibfnamefont {V.}~\bibnamefont {Giesz}},
  \bibinfo {author} {\bibfnamefont {T.}~\bibnamefont {Grange}}, \bibinfo
  {author} {\bibfnamefont {A.}~\bibnamefont {Lema\^itre}}, \bibinfo {author}
  {\bibfnamefont {J.}~\bibnamefont {Demory}}, \bibinfo {author} {\bibfnamefont
  {I.}~\bibnamefont {Sagnes}}, \bibinfo {author} {\bibfnamefont {N.~D.}\
  \bibnamefont {Lanzillotti-Kimura}}, \bibinfo {author} {\bibfnamefont
  {L.}~\bibnamefont {Lanco}}, \bibinfo {author} {\bibfnamefont
  {A.}~\bibnamefont {Auff\`eves}}, \ and\ \bibinfo {author} {\bibfnamefont
  {P.}~\bibnamefont {Senellart}},\ }\href
  {http://dx.doi.org/10.1021/acs.nanolett.5b00876} {\bibfield  {journal}
  {\bibinfo  {journal} {Nano Lett.}\ }\textbf {\bibinfo {volume} {15}},\
  \bibinfo {pages} {6290} (\bibinfo {year} {2015})}\BibitemShut {NoStop}%
\bibitem [{\citenamefont {Galland}\ \emph {et~al.}(2008)\citenamefont
  {Galland}, \citenamefont {H\"ogele}, \citenamefont {T\"ureci},\ and\
  \citenamefont {Imamo\ifmmode~\breve{g}\else \u{g}\fi{}lu}}]{Galland2008}%
  \BibitemOpen
  \bibfield  {author} {\bibinfo {author} {\bibfnamefont {C.}~\bibnamefont
  {Galland}}, \bibinfo {author} {\bibfnamefont {A.}~\bibnamefont {H\"ogele}},
  \bibinfo {author} {\bibfnamefont {H.~E.}\ \bibnamefont {T\"ureci}}, \ and\
  \bibinfo {author} {\bibfnamefont {A.~m.~c.}\ \bibnamefont
  {Imamo\ifmmode~\breve{g}\else \u{g}\fi{}lu}},\ }\href {\doibase
  10.1103/PhysRevLett.101.067402} {\bibfield  {journal} {\bibinfo  {journal}
  {Phys. Rev. Lett.}\ }\textbf {\bibinfo {volume} {101}},\ \bibinfo {pages}
  {067402} (\bibinfo {year} {2008})}\BibitemShut {NoStop}%
\bibitem [{\citenamefont {Vialla}\ \emph {et~al.}(2014)\citenamefont {Vialla},
  \citenamefont {Chassagneux}, \citenamefont {Ferreira}, \citenamefont
  {Roquelet}, \citenamefont {Diederichs}, \citenamefont {Cassabois},
  \citenamefont {Roussignol}, \citenamefont {Lauret},\ and\ \citenamefont
  {Voisin}}]{Vialla2014}%
  \BibitemOpen
  \bibfield  {author} {\bibinfo {author} {\bibfnamefont {F.}~\bibnamefont
  {Vialla}}, \bibinfo {author} {\bibfnamefont {Y.}~\bibnamefont {Chassagneux}},
  \bibinfo {author} {\bibfnamefont {R.}~\bibnamefont {Ferreira}}, \bibinfo
  {author} {\bibfnamefont {C.}~\bibnamefont {Roquelet}}, \bibinfo {author}
  {\bibfnamefont {C.}~\bibnamefont {Diederichs}}, \bibinfo {author}
  {\bibfnamefont {G.}~\bibnamefont {Cassabois}}, \bibinfo {author}
  {\bibfnamefont {P.}~\bibnamefont {Roussignol}}, \bibinfo {author}
  {\bibfnamefont {J.}~\bibnamefont {Lauret}}, \ and\ \bibinfo {author}
  {\bibfnamefont {C.}~\bibnamefont {Voisin}},\ }\href
  {http://link.aps.org/doi/10.1103/PhysRevLett.113.057402} {\bibfield
  {journal} {\bibinfo  {journal} {Phys. Rev. Lett.}\ }\textbf {\bibinfo
  {volume} {113}},\ \bibinfo {pages} {057402} (\bibinfo {year}
  {2014})}\BibitemShut {NoStop}%
\bibitem [{\citenamefont {Ardizzone}\ \emph {et~al.}(2015)\citenamefont
  {Ardizzone}, \citenamefont {Chassagneux}, \citenamefont {Vialla},
  \citenamefont {Delport}, \citenamefont {Delcamp}, \citenamefont {Belabas},
  \citenamefont {Deleporte}, \citenamefont {Roussignol}, \citenamefont
  {Robert-Philip}, \citenamefont {Voisin},\ and\ \citenamefont
  {Lauret}}]{Ardizzone2015}%
  \BibitemOpen
  \bibfield  {author} {\bibinfo {author} {\bibfnamefont {V.}~\bibnamefont
  {Ardizzone}}, \bibinfo {author} {\bibfnamefont {Y.}~\bibnamefont
  {Chassagneux}}, \bibinfo {author} {\bibfnamefont {F.}~\bibnamefont {Vialla}},
  \bibinfo {author} {\bibfnamefont {G.}~\bibnamefont {Delport}}, \bibinfo
  {author} {\bibfnamefont {C.}~\bibnamefont {Delcamp}}, \bibinfo {author}
  {\bibfnamefont {N.}~\bibnamefont {Belabas}}, \bibinfo {author} {\bibfnamefont
  {E.}~\bibnamefont {Deleporte}}, \bibinfo {author} {\bibfnamefont
  {P.}~\bibnamefont {Roussignol}}, \bibinfo {author} {\bibfnamefont
  {I.}~\bibnamefont {Robert-Philip}}, \bibinfo {author} {\bibfnamefont
  {C.}~\bibnamefont {Voisin}}, \ and\ \bibinfo {author} {\bibfnamefont {J.~S.}\
  \bibnamefont {Lauret}},\ }\href
  {http://link.aps.org/doi/10.1103/PhysRevB.91.121410} {\bibfield  {journal}
  {\bibinfo  {journal} {Phys. Rev. B}\ }\textbf {\bibinfo {volume} {91}},\
  \bibinfo {pages} {121410} (\bibinfo {year} {2015})}\BibitemShut {NoStop}%
\bibitem [{\citenamefont {Sarpkaya}\ \emph {et~al.}(2015)\citenamefont
  {Sarpkaya}, \citenamefont {Ahmadi}, \citenamefont {Shepard}, \citenamefont
  {Mistry}, \citenamefont {Blackburn},\ and\ \citenamefont
  {Strauf}}]{Sarpkaya2015}%
  \BibitemOpen
  \bibfield  {author} {\bibinfo {author} {\bibfnamefont {I.}~\bibnamefont
  {Sarpkaya}}, \bibinfo {author} {\bibfnamefont {E.~D.}\ \bibnamefont
  {Ahmadi}}, \bibinfo {author} {\bibfnamefont {G.~D.}\ \bibnamefont {Shepard}},
  \bibinfo {author} {\bibfnamefont {K.~S.}\ \bibnamefont {Mistry}}, \bibinfo
  {author} {\bibfnamefont {J.~L.}\ \bibnamefont {Blackburn}}, \ and\ \bibinfo
  {author} {\bibfnamefont {S.}~\bibnamefont {Strauf}},\ }\href
  {http://dx.doi.org/10.1021/acsnano.5b01997} {\bibfield  {journal} {\bibinfo
  {journal} {ACS Nano}\ }\textbf {\bibinfo {volume} {9}},\ \bibinfo {pages}
  {6383} (\bibinfo {year} {2015})}\BibitemShut {NoStop}%
\bibitem [{\citenamefont {Krummheuer}\ \emph {et~al.}(2002)\citenamefont
  {Krummheuer}, \citenamefont {Axt},\ and\ \citenamefont
  {Kuhn}}]{Krummheuer2002}%
  \BibitemOpen
  \bibfield  {author} {\bibinfo {author} {\bibfnamefont {B.}~\bibnamefont
  {Krummheuer}}, \bibinfo {author} {\bibfnamefont {V.~M.}\ \bibnamefont {Axt}},
  \ and\ \bibinfo {author} {\bibfnamefont {T.}~\bibnamefont {Kuhn}},\ }\href
  {\doibase 10.1103/PhysRevB.65.195313} {\bibfield  {journal} {\bibinfo
  {journal} {Phys. Rev. B}\ }\textbf {\bibinfo {volume} {65}},\ \bibinfo
  {pages} {195313} (\bibinfo {year} {2002})}\BibitemShut {NoStop}%
\bibitem [{\citenamefont {Leggett}\ \emph {et~al.}(1987)\citenamefont
  {Leggett}, \citenamefont {Chakravarty}, \citenamefont {Dorsey}, \citenamefont
  {Fisher}, \citenamefont {Anupam},\ and\ \citenamefont
  {Zwerger}}]{Leggett1987}%
  \BibitemOpen
  \bibfield  {author} {\bibinfo {author} {\bibfnamefont {A.~J.}\ \bibnamefont
  {Leggett}}, \bibinfo {author} {\bibfnamefont {S.}~\bibnamefont
  {Chakravarty}}, \bibinfo {author} {\bibfnamefont {A.~T.}\ \bibnamefont
  {Dorsey}}, \bibinfo {author} {\bibfnamefont {M.~P.~A.}\ \bibnamefont
  {Fisher}}, \bibinfo {author} {\bibnamefont {Anupam}}, \ and\ \bibinfo
  {author} {\bibfnamefont {W.}~\bibnamefont {Zwerger}},\ }\href@noop {}
  {\bibfield  {journal} {\bibinfo  {journal} {Rev. Mod. Phys.}\ }\textbf
  {\bibinfo {volume} {59}},\ \bibinfo {pages} {1} (\bibinfo {year}
  {1987})}\BibitemShut {NoStop}%
\bibitem [{\citenamefont {Wilson-Rae}\ and\ \citenamefont
  {Imamoğlu}(2002)}]{Wilson-Rae2002}%
  \BibitemOpen
  \bibfield  {author} {\bibinfo {author} {\bibfnamefont {I.}~\bibnamefont
  {Wilson-Rae}}\ and\ \bibinfo {author} {\bibfnamefont {A.}~\bibnamefont
  {Imamoğlu}},\ }\href {http://dx.doi.org/10.1103/PhysRevB.65.235311}
  {\bibfield  {journal} {\bibinfo  {journal} {Physical Review B}\ }\textbf
  {\bibinfo {volume} {65}} (\bibinfo {year} {2002})}\BibitemShut {NoStop}%
\bibitem [{\citenamefont {Chassagneux}\ \emph {et~al.}(2017)\citenamefont
  {Chassagneux}, \citenamefont {Jeantet}, \citenamefont {Claude},\ and\
  \citenamefont {Voisin}}]{theorypaper}%
  \BibitemOpen
  \bibfield  {author} {\bibinfo {author} {\bibfnamefont {Y.}~\bibnamefont
  {Chassagneux}}, \bibinfo {author} {\bibfnamefont {A.}~\bibnamefont
  {Jeantet}}, \bibinfo {author} {\bibfnamefont {T.}~\bibnamefont {Claude}}, \
  and\ \bibinfo {author} {\bibfnamefont {C.}~\bibnamefont {Voisin}},\ }\href
  {http://arxiv.org/abs/1706.05204} {\bibfield  {journal} {\bibinfo  {journal}
  {http://arxiv.org/abs/1706.05204}\ } (\bibinfo {year} {2017})}\BibitemShut
  {NoStop}%
\bibitem [{\citenamefont {Nguyen}\ \emph {et~al.}(2011)\citenamefont {Nguyen},
  \citenamefont {Voisin}, \citenamefont {Roussignol}, \citenamefont {Roquelet},
  \citenamefont {Lauret},\ and\ \citenamefont {Cassabois}}]{Nguyen2011a}%
  \BibitemOpen
  \bibfield  {author} {\bibinfo {author} {\bibfnamefont {D.~T.}\ \bibnamefont
  {Nguyen}}, \bibinfo {author} {\bibfnamefont {C.}~\bibnamefont {Voisin}},
  \bibinfo {author} {\bibfnamefont {P.}~\bibnamefont {Roussignol}}, \bibinfo
  {author} {\bibfnamefont {C.}~\bibnamefont {Roquelet}}, \bibinfo {author}
  {\bibfnamefont {J.~S.}\ \bibnamefont {Lauret}}, \ and\ \bibinfo {author}
  {\bibfnamefont {G.}~\bibnamefont {Cassabois}},\ }\href {\doibase
  10.1103/PhysRevB.84.115463} {\bibfield  {journal} {\bibinfo  {journal} {Phys.
  Rev. B}\ }\textbf {\bibinfo {volume} {84}},\ \bibinfo {pages} {115463}
  (\bibinfo {year} {2011})}\BibitemShut {NoStop}%
\bibitem [{\citenamefont {Ai}\ \emph {et~al.}(2011)\citenamefont {Ai},
  \citenamefont {Walden-Newman}, \citenamefont {Song}, \citenamefont
  {Kalliakos},\ and\ \citenamefont {Strauf}}]{Ai2011}%
  \BibitemOpen
  \bibfield  {author} {\bibinfo {author} {\bibfnamefont {N.}~\bibnamefont
  {Ai}}, \bibinfo {author} {\bibfnamefont {W.}~\bibnamefont {Walden-Newman}},
  \bibinfo {author} {\bibfnamefont {Q.}~\bibnamefont {Song}}, \bibinfo {author}
  {\bibfnamefont {S.}~\bibnamefont {Kalliakos}}, \ and\ \bibinfo {author}
  {\bibfnamefont {S.}~\bibnamefont {Strauf}},\ }\href
  {http://dx.doi.org/10.1021/nn102885p} {\bibfield  {journal} {\bibinfo
  {journal} {ACS Nano}\ }\textbf {\bibinfo {volume} {5}},\ \bibinfo {pages}
  {2664} (\bibinfo {year} {2011})}\BibitemShut {NoStop}%
\bibitem [{\citenamefont {Hunger}\ \emph {et~al.}(2010)\citenamefont {Hunger},
  \citenamefont {Steinmetz}, \citenamefont {Colombe}, \citenamefont {Deutsch},
  \citenamefont {H\"ansch},\ and\ \citenamefont {Reichel}}]{Hunger2010}%
  \BibitemOpen
  \bibfield  {author} {\bibinfo {author} {\bibfnamefont {D.}~\bibnamefont
  {Hunger}}, \bibinfo {author} {\bibfnamefont {T.}~\bibnamefont {Steinmetz}},
  \bibinfo {author} {\bibfnamefont {Y.}~\bibnamefont {Colombe}}, \bibinfo
  {author} {\bibfnamefont {C.}~\bibnamefont {Deutsch}}, \bibinfo {author}
  {\bibfnamefont {T.~W.}\ \bibnamefont {H\"ansch}}, \ and\ \bibinfo {author}
  {\bibfnamefont {J.}~\bibnamefont {Reichel}},\ }\href
  {http://stacks.iop.org/1367-2630/12/i=6/a=065038} {\bibfield  {journal}
  {\bibinfo  {journal} {New Journal of Physics}\ }\textbf {\bibinfo {volume}
  {12}},\ \bibinfo {pages} {065038} (\bibinfo {year} {2010})}\BibitemShut
  {NoStop}%
\bibitem [{\citenamefont {Henderson}\ and\ \citenamefont
  {Imbusch}(1989)}]{Henderson1989}%
  \BibitemOpen
  \bibfield  {author} {\bibinfo {author} {\bibfnamefont {B.}~\bibnamefont
  {Henderson}}\ and\ \bibinfo {author} {\bibfnamefont {G.~F.}\ \bibnamefont
  {Imbusch}},\ }\href@noop {} {\emph {\bibinfo {title} {Optical spectroscopy of
  inorganic solids}}}\ (\bibinfo  {publisher} {Clarendon Press},\ \bibinfo
  {address} {Oxford},\ \bibinfo {year} {1989})\BibitemShut {NoStop}%
\bibitem [{\citenamefont {Faraon}\ \emph {et~al.}(2011)\citenamefont {Faraon},
  \citenamefont {Barclay}, \citenamefont {Santori}, \citenamefont {Fu},\ and\
  \citenamefont {Beausoleil}}]{Faraon2011}%
  \BibitemOpen
  \bibfield  {author} {\bibinfo {author} {\bibfnamefont {A.}~\bibnamefont
  {Faraon}}, \bibinfo {author} {\bibfnamefont {P.~E.}\ \bibnamefont {Barclay}},
  \bibinfo {author} {\bibfnamefont {C.}~\bibnamefont {Santori}}, \bibinfo
  {author} {\bibfnamefont {K.-M.~C.}\ \bibnamefont {Fu}}, \ and\ \bibinfo
  {author} {\bibfnamefont {R.~G.}\ \bibnamefont {Beausoleil}},\ }\href
  {http://dx.doi.org/10.1038/nphoton.2011.52} {\bibfield  {journal} {\bibinfo
  {journal} {Nat Photon}\ }\textbf {\bibinfo {volume} {5}},\ \bibinfo {pages}
  {301} (\bibinfo {year} {2011})}\BibitemShut {NoStop}%
\bibitem [{\citenamefont {Auff\`eves}\ \emph {et~al.}(2010)\citenamefont
  {Auff\`eves}, \citenamefont {Gerace}, \citenamefont {G\'erard}, \citenamefont
  {Santos}, \citenamefont {Andreani},\ and\ \citenamefont
  {Poizat}}]{Auffeves2010}%
  \BibitemOpen
  \bibfield  {author} {\bibinfo {author} {\bibfnamefont {A.}~\bibnamefont
  {Auff\`eves}}, \bibinfo {author} {\bibfnamefont {D.}~\bibnamefont {Gerace}},
  \bibinfo {author} {\bibfnamefont {J.-M.}\ \bibnamefont {G\'erard}}, \bibinfo
  {author} {\bibfnamefont {M.~F.}\ \bibnamefont {Santos}}, \bibinfo {author}
  {\bibfnamefont {L.~C.}\ \bibnamefont {Andreani}}, \ and\ \bibinfo {author}
  {\bibfnamefont {J.-P.}\ \bibnamefont {Poizat}},\ }\href {\doibase
  10.1103/PhysRevB.81.245419} {\bibfield  {journal} {\bibinfo  {journal} {Phys.
  Rev. B}\ }\textbf {\bibinfo {volume} {81}},\ \bibinfo {pages} {245419}
  (\bibinfo {year} {2010})}\BibitemShut {NoStop}%
\bibitem [{\citenamefont {Zakharko}\ \emph {et~al.}(2016)\citenamefont
  {Zakharko}, \citenamefont {Graf}, \citenamefont {Schiessl}, \citenamefont
  {H\"ahnlein}, \citenamefont {Pezoldt}, \citenamefont {Gather},\ and\
  \citenamefont {Zaumseil}}]{Zakharko2016}%
  \BibitemOpen
  \bibfield  {author} {\bibinfo {author} {\bibfnamefont {Y.}~\bibnamefont
  {Zakharko}}, \bibinfo {author} {\bibfnamefont {A.}~\bibnamefont {Graf}},
  \bibinfo {author} {\bibfnamefont {S.~P.}\ \bibnamefont {Schiessl}}, \bibinfo
  {author} {\bibfnamefont {B.}~\bibnamefont {H\"ahnlein}}, \bibinfo {author}
  {\bibfnamefont {J.}~\bibnamefont {Pezoldt}}, \bibinfo {author} {\bibfnamefont
  {M.~C.}\ \bibnamefont {Gather}}, \ and\ \bibinfo {author} {\bibfnamefont
  {J.}~\bibnamefont {Zaumseil}},\ }\href {\doibase
  10.1021/acs.nanolett.6b00827} {\bibfield  {journal} {\bibinfo  {journal}
  {Nano Lett.}\ }\textbf {\bibinfo {volume} {16}},\ \bibinfo {pages}
  {3278–3284} (\bibinfo {year} {2016})}\BibitemShut {NoStop}%
\bibitem [{\citenamefont {Ma}\ \emph {et~al.}(2015)\citenamefont {Ma},
  \citenamefont {Hartmann}, \citenamefont {Baldwin}, \citenamefont {Doorn},\
  and\ \citenamefont {Htoon}}]{Ma2015a}%
  \BibitemOpen
  \bibfield  {author} {\bibinfo {author} {\bibfnamefont {X.}~\bibnamefont
  {Ma}}, \bibinfo {author} {\bibfnamefont {N.~F.}\ \bibnamefont {Hartmann}},
  \bibinfo {author} {\bibfnamefont {J.~K.~S.}\ \bibnamefont {Baldwin}},
  \bibinfo {author} {\bibfnamefont {S.~K.}\ \bibnamefont {Doorn}}, \ and\
  \bibinfo {author} {\bibfnamefont {H.}~\bibnamefont {Htoon}},\ }\href
  {http://dx.doi.org/10.1038/nnano.2015.136} {\bibfield  {journal} {\bibinfo
  {journal} {Nat Nano}\ }\textbf {\bibinfo {volume} {10}},\ \bibinfo {pages}
  {671} (\bibinfo {year} {2015})}\BibitemShut {NoStop}%
\bibitem [{\citenamefont {Grange}\ \emph {et~al.}(2015)\citenamefont {Grange},
  \citenamefont {Horneckern}, \citenamefont {Hunger}, \citenamefont {Poiza},
  \citenamefont {Gerard}, \citenamefont {Senellart},\ and\ \citenamefont
  {Auffeves}}]{Grange2015}%
  \BibitemOpen
  \bibfield  {author} {\bibinfo {author} {\bibfnamefont {T.}~\bibnamefont
  {Grange}}, \bibinfo {author} {\bibfnamefont {G.}~\bibnamefont {Horneckern}},
  \bibinfo {author} {\bibfnamefont {D.}~\bibnamefont {Hunger}}, \bibinfo
  {author} {\bibfnamefont {J.-P.}\ \bibnamefont {Poiza}}, \bibinfo {author}
  {\bibfnamefont {J.-M.}\ \bibnamefont {Gerard}}, \bibinfo {author}
  {\bibfnamefont {P.}~\bibnamefont {Senellart}}, \ and\ \bibinfo {author}
  {\bibfnamefont {A.}~\bibnamefont {Auffeves}},\ }\href {\doibase
  arXiv:1501.00931} {\bibfield  {journal} {\bibinfo  {journal} {Phys. Rev.
  Lett}\ }\textbf {\bibinfo {volume} {114}},\ \bibinfo {pages} {193601}
  (\bibinfo {year} {2015})}\BibitemShut {NoStop}%
\end{thebibliography}%

\end{document}